# A model for a plasma ball

Yuri Kornyushin

*Maître Jean Brunschvig Research Unit, Chalet Shalva, Randogne, CH-3975*

A detailed simple model is applied to study a high temperature plasma ball. It is assumed that the ions and delocalized electrons are distributed randomly throughout the charged plasma ball (extra/missing charge is assumed to be found in a thin layer on the surface of a ball). The energy of the microscopic electrostatic field around the ions is taken into account and calculated. It is shown in the framework of this model that regarded charged plasma ball can be stable as a metastable state, when subjected to an external (atmospheric) pressure. Equilibrium radius of a ball is calculated.

## 1. Electrostatic energy of a separate ion

Let us consider first an electrically neutral plasma ball of a volume $V = 4\pi R^3/3$ ($R$ is the radius of a ball), consisting of $n$ ions and $zn$ delocalized electrons [1]. We consider here the ions as point charges, and the delocalized electrons like a negatively charged gas. Let us consider high temperatures, when the plasma is classical [1]. Classical delocalized electrons screen long-range electrostatic field of point charges. The screening Debye-Huckel radius is as follows [2]:

$$1/g = (kTV/4\pi zne^2)^{1/2} = 0.577 R^{3/2}(kT/zn)^{1/2}/e = R^{3/2}/R_0^{1/2}. \qquad (1)$$

where $e > 0$ is elementary charge and $R_0 = 3zne^2/kT$.

The electrostatic field around a separate positive ion submerged into the gas of classical electrons is as follows [2]:

$$\varphi = (ze/r)\exp{-gr}, \qquad (2)$$

where $r$ is the distance from the center of the ion.

The electrostatic energy of this field is the integral over the ball volume of its gradient in square, divided by $8\pi$. The lower limit of the integral on $r$ should be taken as $r_0$, a very small value. Otherwise the integral diverges. Taking into account the extreme smallness of $r_0$ and that the value of $(R_0/R)^{1/2}$ is usually very large, we have the following expression for the electrostatic energy of a separate ion:

$$W_0 = (z^2 e^2/2r_0) - (z^2 e^2 R_0^{1/2}/R^{3/2}). \qquad (3)$$

It is worthwhile to note that the expansion of a ball leads to the decrease in the delocalized electron density. This leads to the increase in the screening radius [see Eq. (1)]. The electrostatic energy of a separate ion increases concomitantly. One can see it, regarding Eq. (3).

## 2. Enthalpy of a charged plasma ball

We regard the ions of the considered plasma ball as randomly distributed. It is well known since 1967, that the electrostatic energy of the randomly distributed ions is just $W = nW_0$ [3].

Let the extra charge be $eN$ (here $N$ is the positive/negative number of the missing/extra electrons). This charge is assumed to be situated on the surface of a ball (as the charges of the same sign repel each other). It produces electrostatic field outside the ball, $\varphi = (eN/r)$ [4]. Inside the ball this charge produces no field [4]. The electrostatic energy of the field of the regarded here extra/missing electrons charge is as follows [4]:

$$U = e^2 N^2 / 2R. \tag{4}$$

The enthalpy of the system regarded is as follows:

$$H(R) = (4\pi/3)PR^3 + (e^2 N^2 / 2R) + (z^2 e^2 n / 2r_0) - (z^2 e^2 n R_0^{1/2} / R^{3/2}), \tag{5}$$

where $P$ is external (atmospheric) pressure.

For a neutral plasma ball, when $N = 0$, $H(R)$ is a monotonic function of $R$. It increases monotonically with the increase in $R$. It has neither a minimum nor a maximum. So, the neutral plasma ball is not stable.

For a charged plasma ball when $P = 0$, $H(R)$ as a function of $R$ has a maximum only. One can see it, regarding Eq. (5). So at a zero external pressure the regarded plasma ball is not stable either.

At non-zero pressure $P$ the enthalpy $H(R)$ may increase with the increase of $R$, may reach a maximum at some $R = R_{max}$, then it may reach a minimum at some $R = R_{min}$ (for a certain range of parameters), and then increases indefinitely. Minimum of $H(R)$ corresponds to a metastable equilibrium. At $P = 0$ the enthalpy $H(R)$ never has a minimum, it has a maximum only.

The system could collapse beyond the minimum, going to some state with a radius $R$, smaller than $R_{min}$. This radius could be so small that the system will be no more a classical one. We shall not discuss this issue here. Anyway the external (atmospheric) pressure may make the regarded plasma ball stable (for some values of parameters).

In classical statistics the kinetic energy is proportional to the number of constituting particles. Per one particle it is equal to $1.5kT$. The contribution of the entropy term is no larger than that [1]. Only this contribution is negative. So these two terms of the opposite signs in the Gibbs free energy could be neglected [1]. So the enthalpy is a main contribution to the Gibbs free energy in the regarded case.

Eq. (5) yields the following equation for the extremal values of $R$, $R_e$:

$$(\partial H / \partial R)_{R = Re} = 4\pi P R_e^2 - (e^2 N^2 / 2R_e^2) + 1.5(z^2 e^2 n R_0^{1/2} / R_e^{5/2}) = 0. \tag{6}$$

Let us consider first the ball in a vacuum ($P = 0$). At $P = 0$ Eq. (6) yields:

$$R_{max} = 9z^4 n^2 R_0 / N^4. \tag{7}$$

For $z = 1$, $n = 10^{21}$, $N = 10^{14}$, $kT = 8.32 \times 10^{-12}$ erg ($T = 60346$ K) we have $R_0 = 8.308 \times 10^{13}$ cm and $R_{max} = 7.477$ cm. This value of $R_{max}$ corresponds to the maximum of the Gibbs free energy.

For a charged plasma ball for the same parameters and $P = 10^{-4}$ bar (a sort of a vacuum) according to Eq. (5) $H(R)$ has a maximum, $H_{max} = 5.152$ J, at $R = R_{max} = 7.53$ cm. Then $H(R)$ has a minimum, $H_{min} = 3.581$ J, at $R_{min} = 25.4$ cm. The calculated energy releases in case of disintegration of a charged plasma ball. The barrier is $H_{max} - H_{min} = 1.571$ J.

For a charged plasma ball in an atmospheric regular pressure, $P = 1$ bar, with $N = 10^{16}$, $n = 10^{20}$, average $z = 2.512$, and $T = 151589$ K we have the same (as in the previous example) values of the maximum and minimum radii, $R_{max} = 7.53$ cm and $R_{min} = 25.4$ cm, but the values of the enthalpy is $10^4$ times larger. The released energy is $H_{min} = 35.81$ kJ and the barrier is $H_{max} - H_{min} = 15.71$ kJ.

## 3. Discussion

When there is no external pressure, increase in the *R* value leads to the decrease in the electrostatic energy of a charge of a ball and to the increase in the electrostatic energy of the ions. This competition could be expected yielding the equilibrium value of a radius, $R_e$. But it is not so, this competition yields only a maximum of *H*(*R*).

When external (atmospheric) pressure is not zero, it seems that the presence of the $4\pi PR_e^2$ term in Eq. (6) should yield the value of the equilibrium radius (additional work is required to expand the ball). It happens, but not at the arbitrary values of the parameters.

Charged plasma ball could be found stable not only at a regular atmospheric pressure, but at a rather low one also (in the framework of the model proposed).